# Sizing and Placement of Battery Energy Storage Systems and Wind Turbines by Minimizing Costs and System Losses

Bahman Khaki, Pritam Das, *Senior Member, IEEE*

*Abstract*— Probabilistic and intermittent output power of wind turbines (WT) is one major inconsistency of WTs. Battery Energy Storage Systems (BESSs) are a suitable solution to mitigate this intermittency which use to smoothen the output power injected to the grid by such intermittent sources. This paper proposes a new optimization formulation using genetic algorithm to simultaneous sizing and placement of BESSs and WTs which result in finding best location and size (capacity) of WTs and BESSs in power system by minimizing total system loss (active and reactive loss) and Costs of WTs and BESSs which improves demand bus voltage profiles. The result of optimization problem is best buses to locate WTs and BESSs and the size (installable active and reactive power) of them. The case studies performed on IEEE 33 bus system, validates the suitability of the formulation for loss minimization and bus voltage profiles improvement in the test system in presence of WT and BESS.

*Index Terms*—Energy Storage Systems, Batteries, Optimal Placement, Optimal Sizing, Wind Turbines, Genetic Algorithm.

## I. INTRODUCTION

Traditionally Energy Storage Systems (ESS) are implemented in power systems to stabilize and compensate local power instabilities in the system. According to standards reactive power support is necessary in power systems for security and operation of the system in presence of renewable energy sources like wind farms. In these standards, there are some performance requirements for WT that besides generating active power they should also contribute in generation of proper amount of reactive power to prevent security problems and voltage instability. Strategically sized and located BESS can assist WTs in meeting these standard requirements.

Different methods of optimization have been reported in the literature. A cost benefit analysis based objective function in distribution system with high penetration Photovoltaic (PV) introduced in [1] by a physical battery model and voltage regulation and peak load shaving oriented energy management system for sizing of energy storage systems (ESS). The graphs in this papers shows that with more PV penetration, more ESS need to be install. Authors in [2] proposes a stochastic cost-benefit analysis model according to wind speed data and use it for sizing of ESS. The results show that installing ESS in power system is not justifiable because of high ESS installation cost and low charging and discharging efficiency. A heuristic procedure based on voltage sensitivity analysis is proposed to find the best location of ESS and a multi-period optimal power flow framework chose to formulate the sizing problem in [3], for which convex relaxations based on semidefinite programming used to solve the problem with objective of average network loss, number of ESS and total installed capacity. There are some papers about sizing of BESS used in grid connected PV system like [4] in which the objective is to minimize the cost associated with net power purchase from the electric grid and the battery capacity loss while at the same time satisfying the load and reducing the peak electricity purchase from the grid. Liu et al. [5] suggested a method to optimally size the ESS so that the wind penetration level can be increased without violating the grid frequency deviation limit. This method did not require dynamic simulation as it was based on theoretical analysis where the proposed method calculated the power spectrum density of the wind fluctuation to achieve time-frequency transformation. In [6] an algorithm based on long-term wind power time series (WPTS) and the calculation of mean wind power was suggested to evaluate the performance of ESS in minimizing the system cost and losses where the charging and discharging cycles of ESS were taken into consideration. The author in [7] proposes to discrete Fourier transform to decompose the required balancing power into different time-varying periodic components, i.e., intraweek, intraday, intra hour, and real-time. Each component can be used to quantify the maximum energy storage requirement for different types of energy storage. This requirement is the physical limit that could be theoretically accommodated by a power system. It is stated that The actual energy storage capacity can be further quantified within this limit by the cost-benefit analysis as future work. Employing linear function like in linear programming model has been proposed in some papers for ESS allocation where in [8] for e.g. the non-linear AC optimal power flow was transformed with a Linear Programming problem, which was then solved using forward backward sweep optimal power flow. Mixed integer linear programming (MILP) was applied in a number of papers for sizing and placement of ESS [9-14]. In [9], the optimal capacity of an ESS in a micro-grid (MG) was computed by minimizing the total MG cost that combined the ESS investment cost and MG operating cost. The formulation of ESS sizing was done using MILP while a Monte Carlo simulation was used to account for random uncertainties during the optimization process, as well as to determine the MG reliability.

Unlike the analytical and numerical approaches for ESS sizing as stated above, the artificial intelligence (AI) approach does not require complex calculation as well as complicated mathematical models and algorithms to obtain the optimal ESS allocation. Instead, the AI approach is to search the solution space. The AI approach does not guarantee the optimal solution, but the solution is generally satisfactory depending on the solution searching ability of that particular AI algorithm. Four methods namely simple, fuzzy, simple and advanced artificial neural network (ANN) were assessed and compared in [15] for the sizing and operation of zinc-bromine flow battery-based ESS in wind power application. The purpose of this work was to minimize the cost associated with the rated power and energy capacity. A new heuristic algorithm, mimicking the improvisation of music players, has been developed and named Harmony Search (HS) in [16]. In [17] optimal placement of battery energy storage is obtained by evaluating genetic algorithm for minimizing net present value related to power losses in addition to its best operation during faced different percentage of load levels with specific electricity price for each level. cost benefit of energy storage installation respect to the energy losses cost is optimized and arbitrage benefits of this installation did not considered. A genetic algorithm (GA)-based bi-level optimization method is developed in [18] that reduces the voltage fluctuations caused by PV penetration through deploying BESS among permitted nodes of a distribution system. Carpinelli et al. in [19] proposed the optimal sizing and placement of BESS by employing a GA to search for the solution space while an inner algorithm based on sequential quadratic programming (SQP) was utilized to solve the objective function which aimed to minimize the total cost required to sustain the network. In [20] at the first stage of optimization a GA was applied to search and update for the optimized allocation parameters, while the fitness function with optimized allocation parameters was evaluated in second stage using AC optimal power flow. The objective function comprised cumulative voltage deviation from the reference voltage and the total power losses with an objective function weight assigned for each. In [21], a sensitivity analysis has been used where the buses with greatest impact on the real power losses with respect to the nodal reactive power were chosen as the candidate buses for the installation of the capacitors. The location of capacitors was then decided using GA by choosing the best combination of number of capacitors and the tap position so that the installation cost and the real power losses can be minimized. In this work, the equations to formulate the storage lifetime more accurately were derived to make the solution more practical. Furthermore, particle swarm optimization (PSO) was applied in optimal ESS allocation as well. In [22], a fuzzy PSO was suggested to solve the ESS allocation problem which reduced the power losses in distribution network. A hybrid multi-objective particle swarm optimization (HMOPSO) which combined the PSO with elitist non-dominated sorting GA was suggested in [23] to size and place the ESS in a system with unstable wind production in order to minimize the system cost and voltage deviation. A methodology to optimally allocate the ESS and DG in a radial distribution system was suggested in [24] where the problem was modelled as nonlinear optimization problem and resolved by employing a modified PSO. In this work, instead of active power only, reactive power of ESS and DG was taken into consideration as well during the optimization. In [25], the objective function is defined for minimizing power loss in distribution system for different seasons, but load flow equations and BESS power limit were not included in optimization problem and just active losses were considered. In [26], the objective function minimizes the hourly social cost of BESS. Wind generation and load are modeled probabilistically using actual data and a curve fitting approach. The cost minimization approach is by probabilistic optimal power flow (POPF). This study lacks PQ performance constraints and reactive power sizing. In [27], the average active power stored into the storage unit at each bus, and total budget for BESS was included in optimization problem formulation, but load flow equations are not included and reactive power demand is neglected. In [28], the formulation of a problem that accounts for: (i) the voltage support of storage systems to the grid, (ii) the network losses and (iii) the cost of the energy-flow towards the external grid are proposed. As the formulated problem is mixed integer, non-convex and non-linear, its solution requires the adoption of heuristic techniques. In [29], an optimal location for BESS has to be identified in the system such that the distribution system losses are minimized. The power flow constraints are included, but power bounds and BESS budget is not in the formulations.

In this paper study the combined sizing of multiple BESS and WTs are based on total injected power both real and reactive required by WTs and BESSs that considered as one novelty of this study. Total loss in this study is considered as complex (active and reactive) as well. It should be noted that the sizing optimization problem should have maximum total active and reactive power limits in the system based on total available budget, generation plans, or needed power to satisfy demand, The maximum required power can be set in optimization problem according to the available budget for installing WTs and/or BESSs or by worst case consideration for which the difference between demand and supply is the highest. However, storage systems are used in many power systems beside renewable energy sources, but these papers neglect placement and sizing of renewable energy sources and BESS simultaneously, which is one of main focuses of this study.

power limits are extracted from these standards and included to the constraints of placement and sizing of WTs which is another novelty in this study.

Fig. 1 shows the conceptual model of power system in presence of distributed energy resources (DERs) and battery energy storage. As mentioned ESS can help in peak shaving, stability and security of such a power system, so that the demand side residential or industrial users can benefit from enhanced power quality and voltage profile improvements. Fig. 2 shows typical PQ performance chart of a WTs in standards [30]. It shows that both active and reactive power generations from WTs are necessary for system operators.

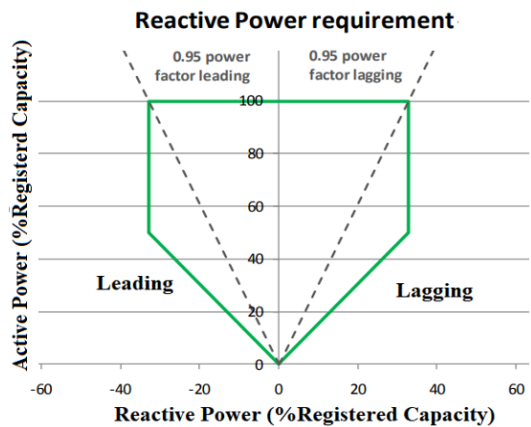

Fig. 2: Typical PQ (performance) chart [30]

The five boundaries (green lines in figure) of this polyhedron are included in constraints of the placement optimization problem in this study. Fig.3 shows the result of load flow analysis at the point of common coupling (PCC) for a typical wind farm. As it is shown in the standard that increasing the quiescent output power of WTs is possible by decreasing the required reactive power support for the WT.

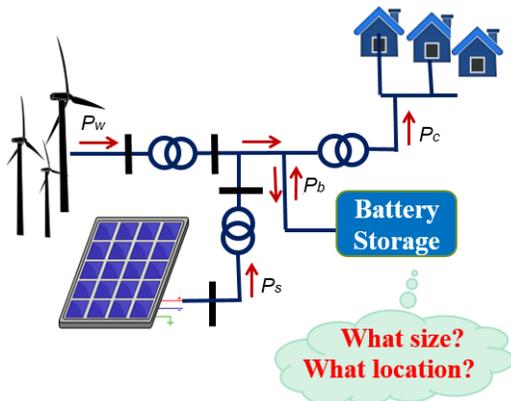

Fig.1: BESS and DERs in system.

Performance requirement of WTs (in lagging and leading reactive power) usually publish by system operators as WT performance PQ chart. In this paper active/reactive

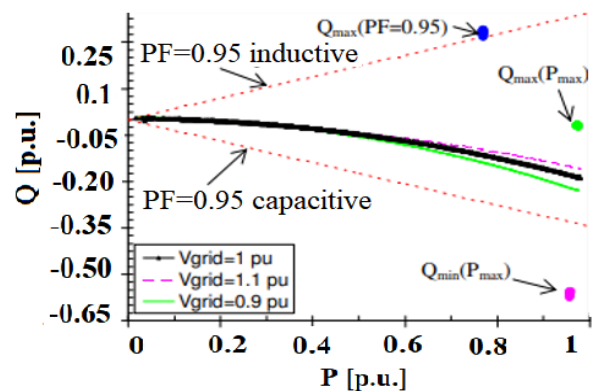

Fig. 3: load flow of typical wind farm at PCC.

WTs can operate near their maximum nominal active power limit (i.e. $P_{max}$), which is one of the benefits of the study presented in this paper. This objective is made possible in this paper by injecting more reactive power from BESSs instead of WTs in simultaneous sizing, so that it can compensate the reactive power support requirement from WTs.

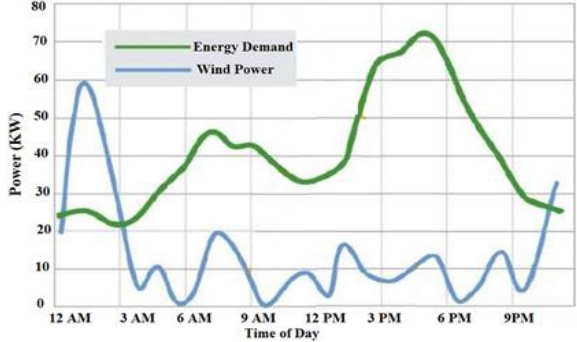

Fig. 4: Typical WT power vs. demand.

This will result in operating renewable energy based plants near their maximum rated output active output power. Fig. 4, shows output power of WT vs. typical demand load. It can be seen that in some periods of a day the demand can be more than power generation by windfarms. Therefore, size of BESSs for worst case can be determined by difference between demand and active power supply in the system.

## II. OPTIMIZATION PROBLEM FORMULATION

Active and reactive power requirements in power system necessitates considering both of these parameters in optimization problem which is not proposed in previous papers [25-30]. Therefore the output parameters of the optimization problem are $P_{BESS}, Q_{BESS}, P_{WT}$, and $P_{WT}$ respectively. System operators' requirements for WTs like reactive power is included in constraints of optimization by typical PQ chart to account for both active and reactive power injection. The result of optimization will show that optimized reactive power compensation from BESS to WTs enhances active power generation from WTs especially in peak hours. For proper placement and sizing of BESSs and WTs, an objective function is defined based on the total loss considering both active and reactive types of the system and should be iteratively calculated. The power flow in this study is based on backward forward load flow (BFLF) that is common for distribution system. This method is formulated base on currents of branches and bus voltages as stated in [31]. For distribution network, the complex load at bus (i), is model as:

$$S_i = (P_i + jQ_i) \quad i = 1, \ldots, N \quad (1)$$

For kth iteration it can be written as:

$$I_i^k = I_i^r(V_i^k) + jI_i^i(V_i^k) = \left(\frac{P_i + jQ_i}{V_i^k}\right)^* \quad (2)$$

Base on load in each branch, and difference between bus voltage, BIBC and BCBV matrixes can be constructed:

$[B]=[BIBC][I]$
$[\Delta V]=[BCBV][B]$ (3)

Where BIBC is the **b**us-**i**njection to **b**ranch-**c**urrent (BIBC) matrix, and BCBV is the **b**ranch-**c**urrent to **b**us-**v**oltage (BCBV) matrix. The solution for distribution load flow can be obtained by solving following equation iteratively:

$$I_i^k = I_i^r(V_i^k) + jI_i^i(V_i^k) = \left(\frac{P_i + jQ_i}{V_i^k}\right)^* \quad (4)$$

$$[\Delta V^{k+1}] = [BCBV][BIBC][I^k] \quad (5)$$

$$[V^{k+1}] = [V^0] + [V^{k+1}] \quad (6)$$

$$V = V_m(\cos\delta + j\sin\delta) \quad (7)$$

$$\delta = tg^{-1}\left(\frac{imag(V)}{real(V)}\right) \quad (8)$$

where $V_i^k, I_i^k$ are the bus voltage and equivalent current injection of the $i^{th}$ bus at the $k^{th}$ iteration, respectively. For loss calculation in the distribution system following equations can be formulated:

$P_L = RI_L^2$      active loss      (9)

$Q_L = XI_L^2$      reactive loss

$P_{TL} = \sum P_L$      active total loss      (10)

$Q_{TL} = \sum Q_L$      reactive total loss      (11)

$S_{TL} = P_{TL} + jQ_{TL}$      total complex loss      (12)

In which $P_L, Q_L, S_L$ are active, reactive and complex loss respectively. The program will calculate the loss in each iteration according to WTs and/or BESSs connected to potential buses. The proposed Genetic Algorithm minimize total loss in the system as described as follows:

$$Minimize \ S_{TL} \quad (13)$$

Subject to

$$[\Delta V] = [BCBV][BIBC][I] \quad (14)$$

$$I_i^k = I_i^r(V_i^k) + jI_i^i(V_i^k) = \left(\frac{P_i + jQ_i}{V_i^k}\right)^* \quad (15)$$

$$[\Delta V^{k+1}] = [BCBV][BIBC][I^k] \quad (16)$$

$$\sum_{i=1}^{n} P_{wt,i} + \sum_{i=1}^{n} P_{BESS,i} = h \quad (17)$$

$$P_{BESS,min} \leq P_{BESS} \leq P_{BESS,max} \quad (18)$$

$$Q_{BESS,min} \leq Q_{BESS} \leq Q_{BESS,max} \quad (19)$$

$$V_{min} \leq V_i \leq V_{max} \quad (20)$$

$$0 \leq P_{wi} \leq P_{WT,max} \quad (21)$$

$$-0.32\ p.u. \leq Q_{wi} \leq 0.32\ p.u. \quad (22)$$

Eqn. (22) is according to Fig. 1(b) standard [30]

$$Q_{wi} - 0.64 P_{wi} \leq 0 \quad (23)$$

Eqn. (23) is active and reactive power inequality (Fig.2) from standard [30].

$$-Q_{wi} - 0.64 P_{wi} \leq 0 \quad (24)$$

Eqn. (24) is active and reactive power inequality (Fig.2) from standard [30].

The final capacity of BESSs and WTs can be derived by solving above optimization problem as:

$$S_{BESSi} = \sqrt{P_{BESSi}^2 + Q_{BESSi}^2} \quad (25)$$

Which is BESS complex power (capacity).

$$S_{WTi} = \sqrt{P_{WTi}^2 + Q_{WTi}^2} \quad (26)$$

Which is wind turbine complex power (capacity).

In equation (17) $h$, stands for total needed power and chosen based on total budget available for BESS and WT installations and/or worst case supply demand requirements. Equation (25) is AC injected power of BESS to grid, thus finally the required/planned DC capacity of BESSs considering the inverter efficiency is:

$$S_{BESS,i}^{DC} = S_{BESS}^{AC}/\eta_i \quad (27)$$

As a second optimization formulation, the cost of BESSs and WTs can be considered in objective function as well. Different kind of batteries being used in power systems and depending on the application (like bulk power system, distributed generation, or as power quality enhancement), the cost of energy storage system is different. In this paper the cost per installable power ($/kW), and per energy ($/kWh) is extracted from reliable references for DG application [33]. The total cost to be minimize in optimization formulation can be define as:

$$cost = \sum_{i=1}^{n} cost_{WT,i} + \sum_{i=1}^{n} cost_{BESS,i} \quad (28)$$

The total cost of energy storage system is actually the cost of storage system ($cost_{SS}$) plus cost of power conversion system ($cost_{PCS}$) and cost of balance of plant (BoP) as follows:

$$cost_{BESS,i} = cost_{SS} + cost_{PCS} + cost_{BoP}$$
$$= c_e \frac{1}{\eta} E + c_p P + cost_{BoP} \quad (29)$$

Table (I) shows the cost related coefficients for Li-ion and Lead-acid batteries which are the majority of battery types use as BESS in power systems. The total cost of BESS in last case study of this paper (considering BESS costs) is extracted from table (I) data.

Table I: cost related data used for different types of batteries

| Battery type | Energy coef. ($c_e$) [$/kWh] | power coef.( $c_p$) [$/kW] | BoP cost [$] | Storage efficiency |
|---|---|---|---|---|
| Li-ion | 500 | 175 | 0 | 85% |
| Lead-acid | 200 | 175 | 50 | 75% |

Table II: wind turbine average cost per watt.

| WT Type | WT size ($\tilde{P}_k$) | Cost [$] ($cost_k$) | Maintenance cost [$] per year |
|---|---|---|---|
| 1 | 1 kW | 2130 | 48 |
| 2 | 1.5 kW | 9000 | 72 |
| 3 | 2.5 kW | 17000 | 120 |
| 4 | 5.0 kW | 32000 | 240 |
| 5 | 10 kW | 64000 | 480 |
| 6 | 15 kW | 100000 | 720 |
| 7 | 800 kW | 1813000 | 38400 |
| 8 | 1.5 MW | 3200000 | 72000 |
| 9 | 2.5 MW | 4014700 | 120000 |

The average cost for some typical size of wind turbines based is calculated and gathered from wind turbine producers as shown in table II, in which average maintenance cost is 48($/kW) according to real data. The objective is to produce highest amount of wind energy at the minimum cost. Therefore, the optimization problem can be written as:

$$Minimize\ \frac{\sum_{i=1}^{n} cost_{WT,i}}{\sum_{i=1}^{n} P_{WT,i}} + \sum_{i=1}^{n} cost_{BESS,i} \quad (30)$$

Subjected to

$$cost_{WT,i} = \sum_{k=1}^{n} n_k \tilde{P}_k\ cost_k \quad (31)$$

$$P_{WT,i} = \sum_{k=1}^{n} n_k \tilde{P}_k \quad (32)$$

In which $n_k$ is integer number of WTs of type $k$, and $i$ is number of buses, and $\tilde{P}$ is nominal output power of each type of WTs (column 1 of table II). Cost of battery storage comprise of cost of energy storage system and cost of power conversion system.

$$\sum_{i=1}^{n} P_i = \sum_{i=1}^{n} P_{WT,i} + \sum_{i=1}^{n} P_{BESS,i} \quad (33)$$

$$cost_{BESS,i} = cost_{ESS} + cost_{PCS} + cost_{BoP}$$

$$= c_e \frac{1}{\eta} E + c_p P + cost_{BoP} \quad (34)$$

The overall objective function considering loss and cost minimization can be written as:

$$Minimize\ w_1 \left[\frac{\sum_{i=1}^{n} cost_{WT,i}}{\sum_{i=1}^{n} P_{WT,i}}\right] +$$
$$w_2 \left[\sum_{i=1}^{n} cost_{BESS,i}\right] + w_3 [S_{TL}] \quad (35)$$

Subject to all constraints and bounds (14-24).

In which $\sum_{j=1}^{3} w_j = 1$, and $w_i$ are weights that can be set according to the part in objective function which is more important to minimize.

The problem formulation tested on IEEE 33 bus system which is a radial distribution system without generation [32].

## III. CASE STUDY AND RESULTS

### A) Only BESS Placement and Sizing

In this part of case study, six cases are considered based on number of BESSs and the optimal location and size calculated for them according to optimization problem stated and genetic algorithm procedure. It is assumed that total 1000 MW of BESS can be install in the system. Table III and Fig. 5 show the result of this part. Result shows addition of BESS as expected with the objective function decreases the total loss and improve the load bus voltages significantly.

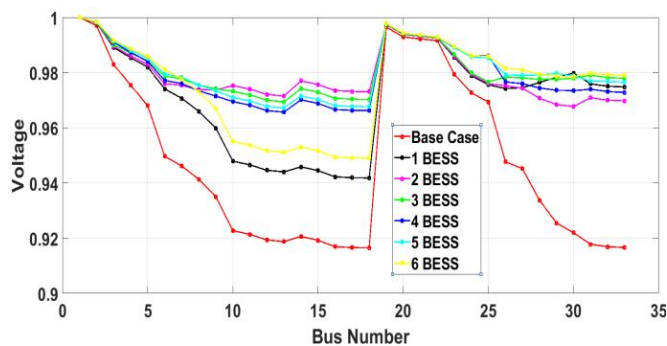

Fig 5: Bus Voltage profiles for placed BESSs

Table III: BESS sizing and placement (6 test cases)

| Case Studies | No. of BESSs | Best BESS Locations (Bus No.) | Size of BESSs ($P_{BESS}+jQ_{BESS}$) | Total Loss ($P_{Loss}+jQ_{Loss}$) | Power loss minimization |
|---|---|---|---|---|---|
| Main Case | 0 | --- | --- | 244 | %0 |
| Case 1 | 1 | 29 | 1000+j1000 | 91.652 | %62.28 |
| Case 2 | 2 | 13 , 30 | 493.22+j546.34<br>506.78+j635.72 | 69.035 | %71.59 |
| Case 3 | 3 | 29, 13, 30 | 314.94+j506.38<br>353.89+j493.29<br>331.17+j489.33 | 63.766 | %73.75 |
| Case 4 | 4 | 13, 29, 24, 30 | 280.50+j484.83<br>250.72+j507.61<br>221.74+j497.81<br>247.03+j477.11 | 66.225 | %72.74 |
| Case 5 | 5 | 28,29,13,24,31 | 200.46+j488.48<br>249.88+j491.75<br>238.71+j498.02<br>148.70+j496.83<br>162.25+j175.79 | 66.243 | %72.73 |
| Case 6 | 6 | 30,25,24,25,29,31 | 172.24+j483.48<br>162.55+j467.20<br>156.77+j500.36<br>161.71+j454.12<br>167.75+j478.55<br>178.98+j0.2933 | 82.798 | %65.92 |

### B) Only WT Placement and Sizing

In this part, six cases considered based on number of WTs to locate and size them optimally according to optimization problem stated and genetic algorithm procedure assuming 1000 MW of installed wind power injection (i.e. $\sum_{i=1}^{n} n_i P_i = h = 1000\ MW$).

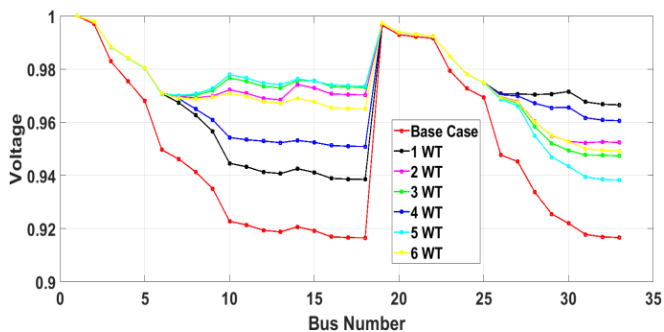

Fig 6: Bus Voltage profiles with WTs placed.

Table IV and Fig. 6 show the results of this part of the study. As we consider reactive constraint for WTs, the result of this case shows less generation of reactive power by WT comparing with first case which was only BESS in system, thus WTs can generate more active power.

Table IV: WT sizing and placement (6 test cases)

| Case Studies | No. of WTs | Best WTs Locations (Bus No.) | Size of WTs $(P_{WT}+jQ_{WT})$ | Total Loss | Power loss minimization |
|---|---|---|---|---|---|
| Main Case | 0 | --- | --- | 244 | %0 |
| Case 1 | 1 | 29 | 1000+j640 | 103.83 | %57.44 |
| Case 2 | 2 | 31, 13 | 198.74+j351.22<br>801.26+j288.76 | 92.12 | %62.24 |
| Case 3 | 3 | 14, 31, 9 | 38.358+j254.69<br>80.771+j279.72<br>880.87+j105.53 | 99.294 | %59.30 |
| Case 4 | 4 | 11, 13, 15, 29 | 18.909+j114.24<br>12.009+j119.85<br>12.219+j128.72<br>956.86+j276.88 | 103.64 | %57.52 |
| Case 5 | 5 | 7, 5, 15, 7, 9 | 1.2699+j68.284<br>4.2502+j207.78<br>0.13718+j106.25<br>2.1906+j44.644<br>992.15+j213.04 | 116.18 | %52.38 |
| Case 6 | 6 | 31, 29, 29, 31, 30, 9 | 7.0981+j61.64<br>2.3244+j291.91<br>3.4859+j102.56<br>3.8872+j68.98<br>2.176+j65.317<br>981.03+j49.578 | 98.916 | %59.46 |

### C) Simultaneous BESS and WT Placement and Sizing for loss minimization

In this part, nine cases considered based on number of WTs and BESS which simultaneously exist in the IEEE 33 bus system, and the best locations and sizes for them are determined based on the total loss minimization of the system. It is assumed that total 1000 MW can be injected by both BESSs and WTs totally. The results of this part shows BESS considerably contribute in both active and reactive power generation to system when the placement and sizing considered simultaneously that help covering both active reactive power requirements of the distribution system. The voltage profile improvement is shown in fig. (7) and Table V shows the optimal solutions using genetic algorithm.

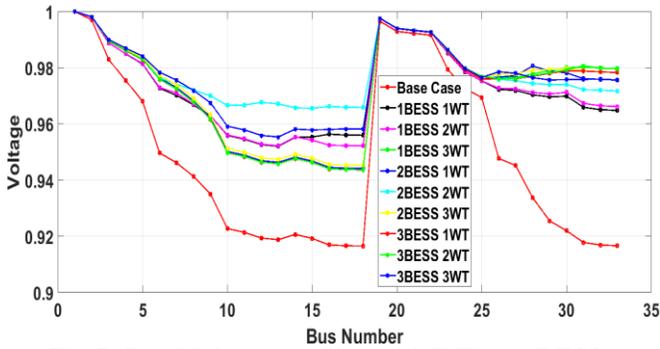

Fig 7: Bus Voltage profiles with WT and BESS placed.

Table V: WT and BESS sizing and placement (9 test cases).

| Case | No. of WT | No. BESS | Best WT place | Best BESS place | Size of WTs $(P_{wt}+jQ_{wt})$ | Size of BESSs $(P_{BESS}+jQ_{BESS})$ | Total loss $(P_{Loss}+jQ_{Los})$ | % loss Min |
|---|---|---|---|---|---|---|---|---|
| Main Case | 0 | 0 | --- | --- | --- | --- | 244 | %0 |
| Case 1 | 1 | 1 | 15 | 29 | 224.96+j143.21 | 775.04+j675.51 | 81.07 | %66 |
| Case 2 | 1 | 2 | 17 | 13, 29 | 32.835+j20.336 | 16.262+j320.83<br>950.90+j524.66 | 89.91 | %63 |
| Case 3 | 1 | 3 | 30 | 29, 25, 31 | 249.15-j0.217 | 252.34+j963.17<br>249.24+j202.89<br>249.27+j216.1 | 83.36 | %65 |
| Case 4 | 2 | 1 | 27, 31 | 29 | 642.62+j319.75<br>184.00+j110.44 | 173.07+j822.29 | 86.54 | %64 |
| Case 5 | 2 | 2 | 15, 11 | 31, 29 | 243.19+j71<br>252.69+j69 | 250.19+j77.664<br>253.94+j998.25 | 64.04 | %73 |
| Case 6 | 2 | 3 | 27, 31 | 30, 29, 25 | 199.27+j46.5<br>207.18+j47.25 | 197.30+j36.788<br>201.74+j999.06<br>194.52+j232.51 | 82.83 | %66 |
| Case 7 | 3 | 1 | 25, 30, 31 | 29 | 257.56+j74.5<br>248.06+j68.75<br>243.06+j49.632 | 251.31+j999.87 | 85.20 | %65 |
| Case 8 | 3 | 2 | 25, 24, 30 | 25, 29 | 46.428+j42.967<br>14.933+j55.717<br>812.30+j50.441 | 125.63+j112.57<br>0.7132+j1000.0 | 85.16 | %65 |
| Case 9 | 3 | 3 | 30, 16, 31 | 31, 29, 25 | 153.28+j45.515<br>189.39+j76.63<br>166.86+j54.175 | 151.17+j74.075<br>183.57+j623.06<br>155.74+j589.27 | 65.79 | %73 |

### D) Optimal BESS and WT placement and sizing considering both costs and total loss

In this part, costs of wind turbine and battery energy storage systems (lithium-ion) added and studied in optimization formulation for placement and sizing of WTs and BESSs simultaneously according to parameters and coefficients mentioned in table I and II for WTs and BESSs. Equation 35 used as objective function and equations 14-24 as constraints. The result in this part is based on Lithium-ion batteries data according to table (I). The voltage profile improvement is shown in fig. 8, and table VI in appendix (A) shows the optimum number of each type of wind turbines base on table II for WT cost minimization, and table VII shows the solution of placement and sizing considering both costs and total loss. According to the results adding costs of WTs and BESSs changed the result of placement and sizing significantly.

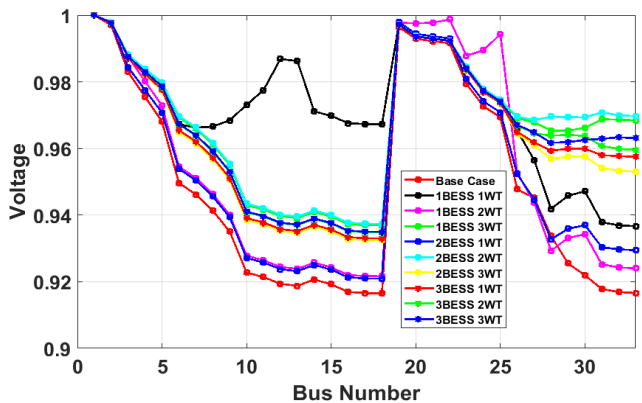

Fig 8: Bus Voltage profiles with WT and BESS placed considering both costs and loss minimization.

IV. CONCLUSION

As energy storage systems are being employed in todays' power systems in presence of renewable energies according to their advantages, the benefits of this study were in simultaneously placement and sizing of WTs and BESS in distribution system considering both active/reactive power injection to grid. Both Cost and total loss minimization investigated in different case studies. Supplying reactive power by WTs is a requirement by system operators nowadays, the proposed solution in this paper was inclusion of typical WT reactive power capability requirements through their PQ chart in optimization problem. BESS can also take a part in injection of reactive power to grid so this study determine the amount of both injected active and reactive power by BESS and WT together to the system by defining different cases and changing number of them in each case. Different capacities for WTs with different costs applied to current studies for cost of WT consideration which is not discussed in previous literature papers, because of lack of cost function for wind turbines output power, thus the method used for WT cost consideration in this paper can be considered as novel procedure based on real market costs for WTs according to table II. Cases defined in four part: only BESSs, only WTs, both BESSs and WTs for loss minimization, and the last one considering both WT and BESS cost and total loss of the power system. Results shows considering BESS placement and sizing in addition to DERs like WTs can help and cover the system active power (peak shaving), and reactive power requirements as well, while results in improved load voltage profiles as in first and third case study sections more. Inclusion of cost in objective function changed the result of placement and sizing of WTs and BESSs significantly.

APPENDIX A

Table VI shows the result of case study (D) in which optimal number of each type of wind turbines base on capacity of WTs for any of 9 case studies is shown according to eqn. (31-32), and Table VII shows the placement and sizing results in case study (D) considering both BESS and WTs costs and total system loss.

Table VI: optimal number of each type of WTs according to eqn. 31-32.

| Case studies | No. | Number of each of 9 types of Wind Turbines according to Table (2) | | | | | | | | |
| Wind Turbine type | | 1 kW | 1.5 kW | 2.5 kW | 5.0 kW | 10 kW | 15 kW | 800 kW | 1.5 MW | 2.5 MW |
| --- | --- | --- | --- | --- | --- | --- | --- | --- | --- | --- |
| No. of WTs (Case1) | WT1 | 1 | 0 | 1 | 1 | 0 | 1 | 0 | 0 | 0 |
| No. of WTs (Case2) | WT1 | 99 | 101 | 226 | 124 | 117 | 107 | 365 | 37 | 0 |
| No. of WTs (Case3) | WT1 | 120 | 170 | 183 | 184 | 176 | 206 | 287 | 33 | 0 |
| No. of WTs (Case4) | WT1 | 858 | 100 | 105 | 103 | 107 | 101 | 97 | 76 | 68 |
| | WT2 | 158 | 207 | 108 | 108 | 101 | 99 | 69 | 39 | 0 |
| No. of WTs (Case5) | WT1 | 99 | 138 | 111 | 194 | 118 | 114 | 103 | 0 | 29 |
| | WT2 | 71 | 61 | 124 | 185 | 106 | 122 | 78 | 0 | 32 |
| No. of WTs (Case6) | WT1 | 100 | 107 | 109 | 101 | 103 | 103 | 92 | 79 | 65 |
| | WT2 | 142 | 158 | 102 | 110 | 104 | 104 | 69 | 38 | 0 |
| No. of WTs (Case7) | WT1 | 105 | 209 | 103 | 99 | 101 | 101 | 89 | 73 | 73 |
| | WT2 | 141 | 164 | 112 | 105 | 109 | 102 | 67 | 41 | 0 |
| | WT3 | 98 | 144 | 103 | 106 | 115 | 104 | 65 | 41 | 0 |
| No. of WTs (Case8) | WT1 | 103 | 183 | 101 | 117 | 126 | 101 | 138 | 200 | 69 |
| | WT2 | 128 | 146 | 117 | 104 | 104 | 101 | 68 | 41 | 0 |
| | WT3 | 102 | 110 | 103 | 115 | 118 | 133 | 67 | 42 | 3 |
| No. of WTs (Case9) | WT1 | 104 | 96 | 104 | 99 | 99 | 99 | 87 | 55 | 0 |
| | WT2 | 65 | 45 | 113 | 107 | 107 | 104 | 74 | 47 | 17 |
| | WT3 | 103 | 106 | 107 | 103 | 103 | 103 | 73 | 47 | 17 |

Table VII: placement and sizing results of case (D)

| Case Studies | No. of WTs | No. of BESSs | Best WT locations | Best BESS locations | Size of WTs | Size of BESSs | WTs Cost | BESSs Cost | Loss |
| --- | --- | --- | --- | --- | --- | --- | --- | --- | --- |
| Main Case | 0 | 0 | --- | --- | --- | --- | --- | --- | 244 |
| Case 1 | 1 | 1 | 6 | 11 | 4.4-j0.3 | 995.5+j299.4 | 1.7e3 | 7.6e5 | 143.8 |
| Case 2 | 1 | 2 | 21 | 24<br>24 | 353+j3.28 | 289.96+j476.41<br>357.01+j542.56 | 8.19e8 | 4.82e5 | 183 |
| Case 3 | 1 | 3 | 27 | 30<br>30<br>31 | 290.3+j31.6 | 208.7+j171.6<br>260.3+j148.9<br>240.5+j137.4 | 6.04e8 | 5.37e5 | 110.7 |
| Case 4 | 2 | 1 | 18<br>31 | 30 | 814.13+j6<br>121.31+j10.2 | 64.5+j61.65 | 1.7e9 | 1.21e4 | 184.1 |
| Case 5 | 2 | 2 | 30<br>29 | 30<br>29 | 287.2+j62.4<br>270.9-j0.19 | 220.15+j265.1<br>221.6+j204.08 | 8.75e8 | 3.37e5 | 108.2 |
| Case 6 | 2 | 3 | 29<br>30 | 30<br>27<br>28 | 666.1+j1.8<br>119+j4.5 | 70.6+j10<br>72.2+j12.1<br>71.1+j10.8 | 1.44e9 | 1.09e5 | 137.3 |
| Case 7 | 3 | 1 | 29<br>30<br>31 | 28 | 684.05-j2.81<br>122.18-j7.3<br>121.5+j11.1 | 72.24+j92.09 | 1.8e9 | 5.51e5 | 133.1 |
| Case 8 | 3 | 2 | 29<br>30<br>29 | 29<br>29 | 621+j21.8<br>122+j10.1<br>132+j13.7 | 63.7+j133.4<br>60.27+j90.7 | 2.47e9 | 9.46e4 | 123.4 |
| Case 9 | 3 | 3 | 31<br>31<br>30 | 31<br>29<br>27 | 215+j3.4<br>180+j2.81<br>178+j5.08 | 133+j108<br>132+j80<br>159+j91 | 1.39e9 | 3.2e5 | 121.5 |